\documentclass[a4paper]{jpconf}
\usepackage{graphicx}
\usepackage{gensymb}
\usepackage{amssymb}
\newcommand{\gs}{\gtrsim}
\renewcommand{\lsim}{\lesssim}

\usepackage{comment}
\begin{document}
\title{SS433 JET PRECESSION: INFRARED DISPERSION DOPPLER EFFECT}

\author{Ericson Lopez}

\address{Quito Astronomical Observatory of National Polytechnic School and  Physics Department, Sciences Faculty of National Polytechnic School, Quito, Ecuador.}
\ead{ericsson.lopez@epn.edu.ec}
\author{Jairo Armijos}
\address{Quito Astronomical Observatory of National Polytechnic School, Quito, Ecuador}
\ead{jairo.armijos@epn.edu.ec}

\begin{abstract}
 In the infrared part of the electromagnetic spectrum, the relativistic dispersion theory is applied to the SS433 jets, to describe its dispersion properties. In particular, we here investigate the dispersion Doppler effect of the radiation in the relativistic cold plasma approximation for this micro-quasar. 

We find that the Doppler plasma line displacements in SS433 are affected by the plasma dispersion only in a narrow frequency band; in the far infrared (far-IR) one. As a consequence, although the shift $Z$ modulation due to the precession of the SS433 jets is well described by previous works in the optical range, it has to be corrected by plasma dispersion effects in the far-IR range.   
\end{abstract}

\section{Introduction}
The micro-quasar SS433, located at a distance of about $5500~ pc$, is a magnificent laboratory  to study the physical mechanics of the jet formation and radiation generation, beside other interesting issues like jet precession  and its stability.\\ 

From the observations, along the length of SS43 jets, three main emitting regions can be resolved: the internal X-ray emitting region at $r \leq 10^{12} ~ cm$, the middle IR and optical at $r \sim  10^{13}$ - $10^{15} ~cm$  and the external radio emitting region at $r\sim 3\times 10^{15}$ - $10^{17} ~ cm $.\\
	  
The radiation coming from the jets is variable and the emission lines change their wavelength with a 164-day period  \cite{ma79a}.  This variability is observed in the high frequency region as well as in the optical frequency one.  Using spectrometers on ASCA, the radiation coming from the X-ray emitting region of the jets was resolved by  \cite{ko98}. These observations demonstrate that the line displacements follow the same Doppler shift patter as that seen in the optical region.  Consequently, it is concluded that the jet speed at the inner regions of ($r\leq 10^{12} ~cm$) has the same mean value of around $0.26~c$. \\

In the present contribution, we evaluate the radiation dispersion process to estimated the periodic Doppler shift of emission lines, detected in optical photometric and spectroscopic measurements  of the SS433 jets, but  extended to the infrared part of electromagnetic spectrum.\\

\section{Far-IR dispersive Doppler effect}
The periodic  $Z$ modulation of the emission lines detected in photometric and spectrometric observation, in the optical part of the electromagnetic spectrum, is well explained by the Margon and Anderson kinematic model \cite{ma79a,ma79b,ma79c},
in what the jet precession is the  essential physical reason to provoke  the periodic  Doppler shift of the lines. \\ 

In the Margon-Anderson model the plasma is considerate isotropic and the observation frequencies are much higher that the plasma frequency ($ w \gg w_p$) , so the plasma refraction is correctly  described by an refraction index equal to unity ($ n = 1$) .  However, for moderated frequencies ( $w \gs w_p$)  the plasma becomes anisotropic and the characteristic refraction index  is less than unity ($ n< 1$), so in the infrared band, corrections must be introduced in the kinematic model to describe the Doppler displacement  ($Z$) of the lines. In this case, the Doppler displacement, considering the dispersion, is describe by the following basic equation:\\

$$ 1+ Z(t, v, \lambda) = \gamma~ [1 \pm n(t, v, \lambda) ~\beta ~\cos \Psi(t)] $$
\noindent
where $Z$ is the Doppler modulation, $t$ the time, $v$ is the flow speed, $\lambda$ is the observation wavelength, $\gamma$ is the Lorentz factor, $n$ the refraction index and  $\beta= v/c$. The dispersion angle $\Psi(t) = \cos(i) \cos (\theta) + \sin(i) \sin(\theta) \cos\phi(t)$, with $i$  the jet axis average inclination respect to the line of  sight ($i = <\phi> = 79\degree$), $\theta$ is the precession half opening angle  ($20\degree$) and $\phi = (2\pi /P) ~ t $. $P = 164 $~days is the precession period of the jet. \\

We assume that the jets are an expanding radiating cooling plasma so that their physical mean parameters like temperature, electron density, magnetic field, among other, are falling down proportional to the radial coordinate $r$. Then, the tensorial elements given by Lopez \cite{lo04,lo06} can be evaluated for the SS433 source.\\

The Figure 1 shows the results of the kinematic model by Margon et al. \cite{ma79a,ma79b,ma79c} that works well for frequencies much higher than the plasma frequency ($w\gg w_p= 5.6\times 10^{10} Hz$). In this optical regimen the refraction index is the unity ($n=1$) and the model is in good agreement with the observations.\\
\begin{figure}[h]
\begin{center}
\includegraphics[width=0.7\linewidth]{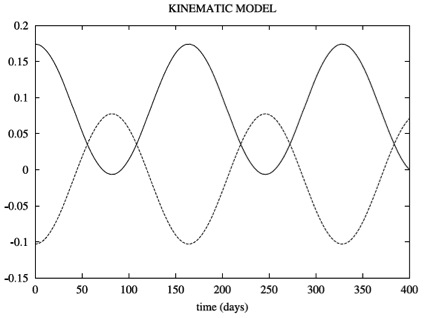}
\end{center}
\caption{Kinematic Model for the optical emitting region of SS433 ($r\sim 10^{14} ~ cm$). Doppler shift $Z$ modulation: Jet precession (full line) and counter-jet precession (dashed line). }
\end{figure}

At the far-infrared frequencies, near to the plasma frequency $w_p = 2.5 \times 10^{12}~Hz $, for what the mean refraction index is about $ 0.7$, the Doppler shift modulation is described by the equation (1), that takes into account the plasma dispersion. Figure 2 describes the Doppler modulation, showing that for the far infrared band, in the inner regions of the jets, the maximum Doppler displacement  $ Z_{red} = 0.135$ and the minimum $Z_{blue}= - 0.006$. The jets, however, still move with the same 164-days precession period, as it was observed in the optical regime. Moreover, we see that the far-infrared curves keep symmetric around $Z=0.04$, value that corresponds to the jet velocity ($v\sim 0.26˜c $),  fully agreement with the quadratic Doppler effect used by Margon et al. \cite{ma79a,ma79b,ma79c} to obtain this value.\\
\begin{figure}[h]
\begin{center}
\includegraphics[width=0.7\linewidth]{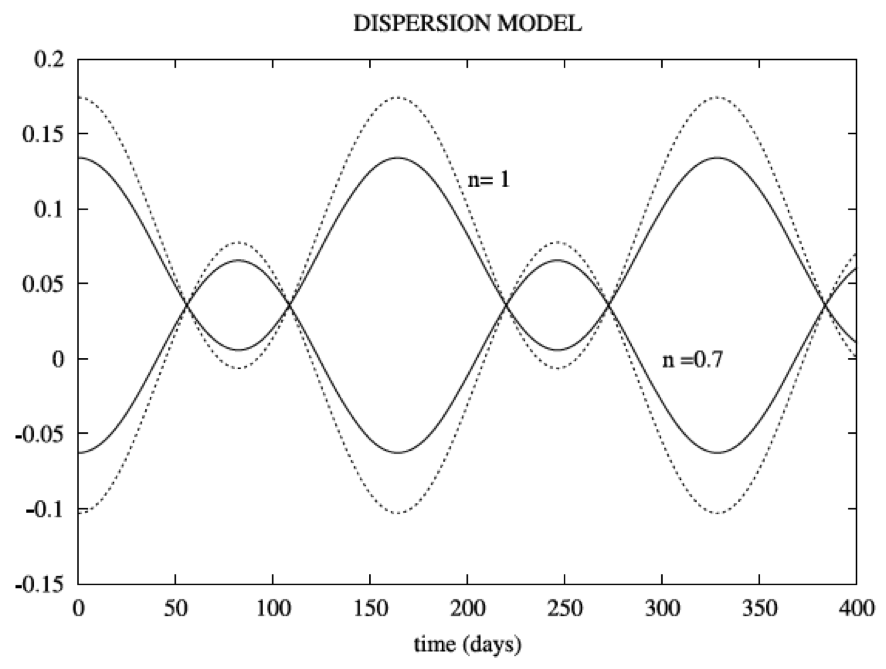}
\end{center}
\caption{Kinematic Dispersion Model for the inner region of SS433 ($r\lsim 10^{12} ~cm$). Doppler shift $Z$ modulation at frequency of $w=2.5 \times 10^{12} ~Hz$: the far-IR jet precession (full line) and the optical jet precession (dashed line).  }
\end{figure}

\section{Discussions and conclusions}
Along the SS433 jet, the rarefied electron plasma shows dispersion features that divide the jet in three well-defined regions: the inner X-ray and the intermediate optical regions which may be considered isotropic media, and the outer radio-emitting region that becomes anisotropic due to the low electron density and the substantial influence of the magnetic field \cite{lo06}. \\

The dispersion curves for the relativistic jets of SS433 also reveal the existence of a narrow range of frequencies where the refraction index is significantly smaller that the unity. For the inner regions of the jet ($r\lsim 10^{12}$), this range falls in the far-IR part of the electromagnetic spectrum ($1.8 \times 10^{12} \lsim w \lsim 2\times 10^{13} ~Hz$) for which the refraction index is 0.7. Consequently, the periodic Doppler shift modulation caused by the jet precession is significantly affected by dispersion effects and the maximum and minimum Doppler shifts decrease to $Z_{red}=0.135$ and $Z_{blue}=-0.006$, respectively, in the far infrared part of  the electromagnetic spectrum. At higher frequencies the radiation is not affected by the dispersion effect, since the relativistic plasma is isotropic and the refraction index is the unity.\\

These dispersion effects should be important to take in consideration observing sources with relativistic jets, at frequencies for what the dispersion properties of the environment is characterized by a refraction index less than the unity.\\

New observational instruments like the ALMA telescope, sensitive to the far infrared emission is able to detect millimeter and sub-millimeter wavelengths in the jets of the SS433 source and could be used to corroborate the predictions made in the current contribution.  \\

\end{document}